\documentclass[twoside]{dis08}
\usepackage[latin1]{inputenc}
\usepackage[dvips]{graphicx,epsfig,color}
\usepackage{wrapfig,rotating}
\usepackage{amssymb,amsmath,array}

\begin{document}
\title{Searches for GMSB at the LHC}

%***********************************************************************
% AUTHORS INFORMATION AREA
%***********************************************************************
\author{Mark Terwort
%
% Optional short acknowledgment: remove next line if non-needed
\thanks{on behalf of the ATLAS and CMS collaborations}
%
% DO NOT MODIFY THE FOLLOWING '\vspace' ARGUMENT
\vspace{.3cm}\\
%
% Addresses and institutions (remove "1- " in case of a single institution)
Universit\"at Hamburg - Institut f\"ur Experimentalphysik \\
Luruper Chaussee 149, D-22761 Hamburg - Germany
%
% Remove the next three lines in case of a single institution
%\vspace{.1cm}\\
%2- School of Second Author - Dept of Second Author \\
%Address of Second Author's school - Country of Second Author's school\\
}
%***********************************************************************
% END OF AUTHORS INFORMATION AREA
%***********************************************************************

\maketitle

\begin{abstract}
In SUSY models with Gauge Mediated Supersymmetry Breaking (GMSB) a gravitino is
the lightest SUSY particle (LSP), while a neutralino or a slepton is the next-to-lightest (NLSP).
For the pair-production of SUSY particles at the LHC large missing
transverse energy (from the gravitinos) and two photons or leptons in the final state are
expected, if the NLSP decays inside the detector. In the case of stable NLSPs,
interesting signatures consist of non-pointing photons or heavy stable charged particles.
In this presentation methods are presented, which were developed by the ATLAS and CMS collaborations.
The expected performance derived from simulations is summarized.
\end{abstract}

\section{Introduction}

\begin{wrapfigure}{r}{0.4\columnwidth}
\centerline{\includegraphics[width=0.35\columnwidth]{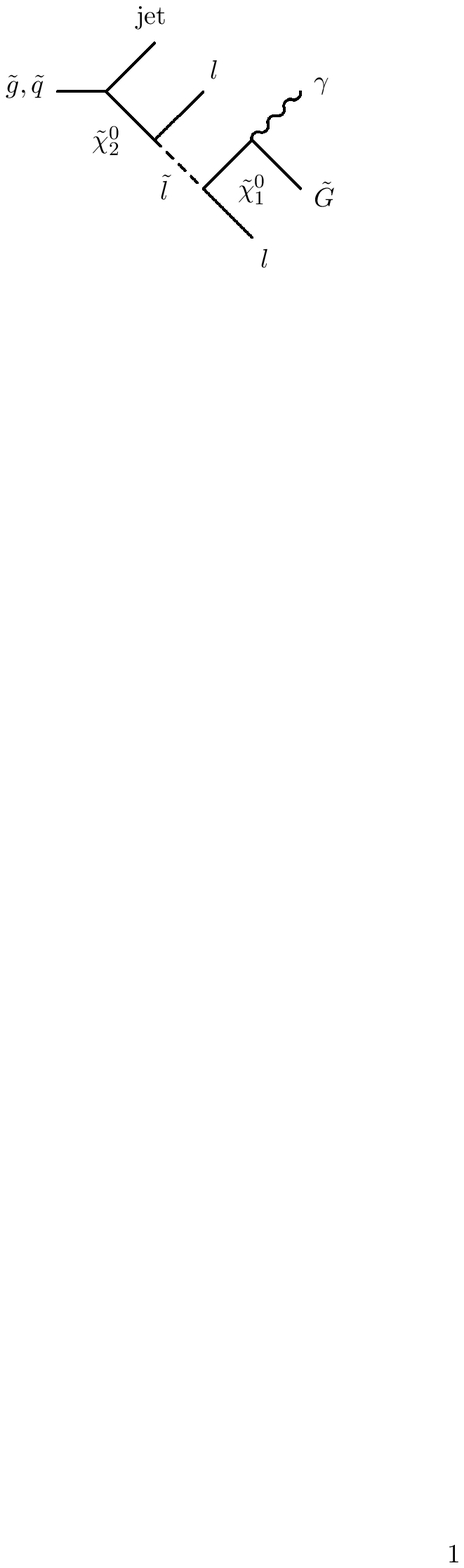}}
\caption{Typical GMSB decay chain for a neutralino NLSP decaying to a photon and a gravitino.}\label{fig:photon:feynman}
\end{wrapfigure}

Since no supersymmetric particles have been found so far, supersymmetry has to be broken, if it exists.
Within Gauge Mediated Supersymmetry Breaking (GMSB) models the breaking in a
secluded sector is mediated via gauge interactions to the electro-weak scale, such that
the superpartners of the standard particles aquire additional masses. If R-parity is
assumed to be conserved, the lightest SUSY particle (LSP), which is the goldstino/gravitino,
is stable. It is nearly massless, since its mass is proportional to the breaking scale,
which usually lies well below the Planck scale. The next-to-lightest SUSY particle (NLSP)
is often either the lightest neutralino or the lightest stau, which decay to the LSP
and their standard model partners with a lifetime also depending on the breaking scale.
The minimal GMSB model is characterised by six fundamental parameters. These are
the effective SUSY breaking scale $\Lambda$, the mass scale of the messengers $M_m$,
the number of messenger SU(5) supermultiplets $N_5$, the ratio of the Higgs vacuum
expectation values $\tan\beta$, the sign of the higgsino mass term $\text{sign}(\mu)$ and the
scale factor of the gravitino coupling $C_{\text{grav}}$, which determines the NLSP lifetime.
Besides many other measurements, discovering signatures of GMSB and measuring the properties
of the NLSP are among the physics goals of the ATLAS and CMS detectors~\cite{url}.
The current status of the Tevatron searches for GMSB is given in~\cite{Tevatron}.
\newpage
\section{Di-photon final states}

\begin{wrapfigure}{r}{0.45\columnwidth}
\centerline{\includegraphics[width=0.45\columnwidth]{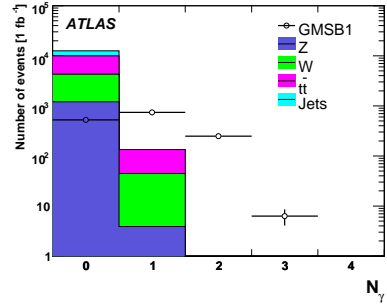}}
\caption{Number of photons after the ATLAS preselection for 1 fb$^{-1}$ for signal and background.}\label{fig:photon:photons}
\end{wrapfigure}

In GMSB models with $N_5 = 1$ and low $\tan\beta$ the lightest neutralino $\tilde{\chi}^0_1$ is the
NLSP which decays to a gravitino and a photon, a $Z^0$ or a Higgs, while the branching ratio for the decay
to photons is usually $> 90\%$. Additionally several jets are expected in the
event, depending on the branching ratios of the squarks and gluinos decaying to various types of SUSY particles.
A typical decay chain is shown in Fig.~\ref{fig:photon:feynman}. If the scale factor of the
gravitino coupling $C_{\text{grav}}$ is of the order of unity, the neutralinos have a short lifetime and
events are expected with two high-energy photons which originate close to the interaction
vertex ({\it prompt photons})\footnote{For larger values of $C_{\text{grav}}$ the photons might be generated somewhere in the inner
detector and do not point back to the interaction vertex ({\it non-pointing photons}). This case is not
discussed here.}.
The corresponding event signatures and the discovery potential of the ATLAS detector for these models
are discussed in~\cite{CSCnote} and are briefly summarized in this section. Similar studies have been performed for
CMS with comparable results.

\subsection{Prompt photon selection}

\begin{wraptable}{r}{0.6\columnwidth}
\centerline{\begin{tabular}{|r||r|r||r|r|r|}
\hline
    N$_{\gamma}$ & Sig &
    $\sum$ BG & N$_{W}$ & N$_{Z}$ & N$_{t\tilde{t}}$ \\ 
    \hline
    0 & 1287.4 & 929.6 & 274.4 & 21.0 & 632.8 \\ 
    1 &  902.9 &  51.7 &  19.5 &  2.0 &  30.1 \\
    2 &  252.9 &   0.1 &   0.0 &  0.0 &   0.1 \\ 
    \hline
\end{tabular}}
\caption{Number of selected signal and background events for 1 fb$^{-1}$ for different cuts on the number
    of photons.}
\label{tab:cutflow}
\end{wraptable}

For detailed studies of the prompt photon signature the ALTAS GMSB1 benchmark point with
$\Lambda=90$ TeV, $M_m=500$ TeV, $N_5 = 1$, $\tan\beta=5$, $\text{sign}(\mu)=+$ and $C_{\text{grav}}=1$ is considered.
For this point the branching ratio of the decay of the lightest neutralino to a photon and a gravitino is $\sim 97\%$,
and the total SUSY production cross section is $\sim 7.8$ pb. 48.9\% (16.4\%) of the signal events
have one (two) photons with $p_T>20$ GeV in the fiducial acceptance region ($|\eta|<2.5$) used for photon identification.
Common identification criteria are used, based on the calorimeter energy deposition and shower shapes.
All relevant Standard Model background processes (mainly QCD jets, gauge boson production with additional jets and $t\bar{t}$ production)
are taken into account using Monte-Carlo events based on a full G4-simulation of the ATLAS detector.

In order to separate this background from the signal a standard preselection
for SUSY-like signatures is performed: At least four jets must be present with $p_T>50$ GeV ($p_T>100$ GeV for the leading jet)
and the missing transverse energy $E_T^{\text{miss}}$ must exceed 100 GeV and 20\% of the effective mass $M_{\text{eff}}$ which is the
scalar sum of $E_T^{\text{miss}}$ and the transverse momenta of the four leading jets.
However, the striking feature of GMSB1 is the large number of hard photons in the final state
which are not present in Standard Model processes. Figure~\ref{fig:photon:photons} shows the number of reconstructed
photons with $p_T>20$ GeV and $|\eta|<2.5$ after the preselection for signal and background. With the additional
requirement of two high energy photons the signal is almost background free. The number of selected signal and
background events for 1 fb$^{-1}$ for different cuts on the number of photons are listed in Tab.~\ref{tab:cutflow}.
Additionally, it has been required that the events pass a combination of single (non-isolated, $p_T>60$ GeV)
and di-photon (isolated, $p_T>20$ GeV) triggers.

\subsection{Discovery potential}

\begin{wrapfigure}{r}{0.45\columnwidth}
\centerline{\includegraphics[width=0.45\columnwidth]{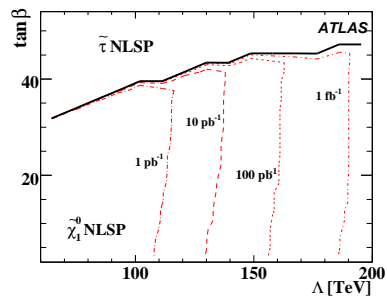}}
\caption{Contour lines with 5 signal events in the $\Lambda$-$\tan\beta$ plane for different integrated luminosities.}\label{fig:photon:scan}
\end{wrapfigure}

In order to estimate the discovery potential of the di-photon channel in a larger part of the parameter space, a scan
of the breaking scale $\Lambda$ and $\tan\beta$ has been performed using a fast simulation. The other parameters have
been fixed at $M_m=500$ TeV, $N_5 = 1$, $\text{sign}(\mu)=+$ and $C_{\text{grav}}=1$.
In this region the neutralino is usually the NLSP, except those with large $\tan\beta$. In these parts other
channels need to be investigated to discover GMSB SUSY, which are discussed in the next section.
Figure~\ref{fig:photon:scan} shows the contour lines with 5 signal events for different integrated luminosities after the
above defined selection. In the regions below and left of the lines a discovery could be made with the corresponding amount
of data assuming the background to be essentially zero. In the high $\tan\beta$ region the $\tilde{\tau}$ is the NLSP and no photons
occur in the decay chain. In general it can be said that the discovery potential is very high in larger regions of the
parameter space than the current limits are excluding, even with early data.

\section{Heavy stable charged particles}
\subsection{Introduction}

\begin{wrapfigure}{r}{0.4\columnwidth}
\centerline{\includegraphics[width=0.35\columnwidth]{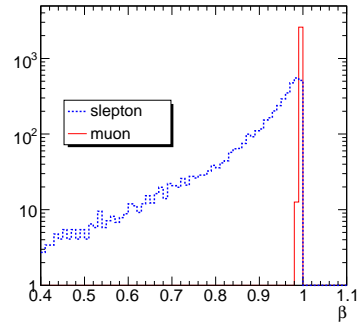}}
\caption{$\beta$ distribution of staus and muons in the ATLAS GMSB5 benchmark point.}\label{fig:HSCP:beta}
\end{wrapfigure}

In regions of the GMSB parameter space where $\tan\beta$ is large, right-handed sleptons are typically the NLSPs,
while the case of a $\tilde{\tau}$ NLSP has been extensively studied by the ATLAS and CMS collaborations.
These particles can have an arbitrarily long lifetime
depending on the gravitino coupling. Here, only the case of quasi-stable staus, which decay outside the detector, is
considered. Since they have a very large mass compared to muons, they traverse the detector
with a velocity $\beta$ which can be smaller than 1. Figure~\ref{fig:HSCP:beta} shows the $\beta$ spectrum of staus
and muons in the ATLAS GMSB5 benchmark point with the parameters $\Lambda=30$ TeV, $M_m=250$ TeV, $N_5=3$, $\tan\beta=5$,
$\text{sign}(\mu)=+$ and $C_{\text{grav}}=5000$. For large $\beta$ these $\tilde{\tau}$s are indistinguishable from
ordinary muons and one can use muon triggers to trigger on these events.

When one tries to identify particles with small $\beta$, one has to pay attention to the dimensions of the detector.
For example, ATLAS extends over 20m in length from the interaction point and since the bunch crossing rate is 40 MHz,
3 events are present in the detector simultanously. For small $\beta$ the stau
could then be assigned to a wrong bunch crossing. However, in GMSB5 99\% of the staus have $\beta>0.7$ and such the trigger efficiency
for a single muon trigger on level 1 with $p_T>40$ GeV is high (95\%)~\cite{CSCnote} for this benchmark point, but might get
worse for larger stau masses.

\subsection{$\beta$ measurement and selection}

In the following a study of CMS on the $\beta$ measurement is summarized~\cite{CMS}.
Due to the low $\beta$ the particles will arrive at the muon system with a delay compared to highly relativistic muons.
Thus, the reconstructed hits for each tube will be shifted with respect to its real position, since the tubes in consecutive
layers are staggered such that a typical track passes alternatively on the left and on the right side of a sensitive wire.
The result is a zig-zag pattern from which the time of flight can be calculated. Another method to measure the velocity
of heavy stable charged particles in CMS uses the higher ionisation of these particles with respect to ordinary MIPs.
Since the CMS tracking detector is able to measure the deposited energy of each hit, $\beta$ can be obtained from $\frac{dE}{dx}$:
\begin{equation}
\beta^{-1}=\sqrt{K\frac{dE}{dx}}, \nonumber
\end{equation}
where the factor $K$ is a constant which can be obtained from data using low momentum protons.
The velocity can be used to reject the background muons
and select a clean signal sample. An analysis strategy using both $\beta$ measurement procedures is demonstrated in~\cite{Loic}.

\section{Conclusions}

GMSB models are a possibility to break SUSY on a low scale compared to generic mSUGRA models. A consequence is
that the LSP is typically the gravitino which couples to the NLSP and its Standard Model partner. Depending on the
nature of the NLSP, interesting signatures containing for example high energy photons or heavy stable charged particles are the result.
Based on simulations both ATLAS and CMS have developed dedicated search strategies for these events, which might allow the
discovery of supersymmetric particles already with early LHC data.

% ****************************************************************************
% BIBLIOGRAPHY AREA
% ****************************************************************************

\begin{footnotesize}
% IF YOU DO NOT USE BIBTEX, USE THE FOLLOWING SAMPLE SCHEME FOR THE REFERENCES
% ----------------------------------------------------------------------------

% ----------------------------------------------------------------------------
\end{footnotesize}


\begin{thebibliography}{99}
% Please replace the numbers for   contribId   and   sessionId
% in the following URL. You can get this information by going to 
% http://indico.cern.ch/confAuthorIndex.py?confId=24657
% and search for your contribution and click on the title
% Be aware: '&amp;' must be replaced by simple '&' as in example below
\bibitem{url} Slides: \\ 
\verb$http://indico.cern.ch/contributionDisplay.py?contribId=77&sessionId=15&confId=24657$
\bibitem{Tevatron} Yurii Maravin, {\it Tevatron Searches for GMSB and RPV SUSY}, these proceedings.
\bibitem{CSCnote} ATLAS collaboration, {\it Studies of SUSY signatures with high-$p_T$ photons or
long-lived heavy particles in ATLAS}, SUSY CSC 8, to be published.
\bibitem{CMS} CMS Collaboration, {\it Search for Heavy Stable Charged Particles with 100 pb$^{-1}$
and 1 fb$^{-1}$ in the CMS experiment}, CMS note, CMS PAS EXO-08-003.
\bibitem{Loic} Loic Quertenmont, {\it LHC Searches for Heavy Stable Charged Particles}, these proceedings.

\end{thebibliography}
\end{document}